\documentclass[preprint,prd,showpacs,preprintnumbers,amsmath,amssymb,floatfix]{revtex4-2}

\usepackage{amsmath}
\usepackage{graphicx}% Include figure files
\usepackage{dcolumn}% Align table columns on decimal point
\usepackage{bm}% bold math
\usepackage{ulem} %% for strike-through
\usepackage[usenames]{color}
\usepackage{epstopdf}
\usepackage{float}
\usepackage{multirow}

\usepackage{array}
\usepackage{booktabs}   % 用于美化表格线条
\usepackage{subcaption} % 更现代、功能更强的子图宏包，用于替换 subfigure
\usepackage{tabularx}

\newcommand{\nc}{\newcommand}       % new command
\nc{\vc}[1] {\mbox{\boldmath $#1$}} % boldmath(vector)
\nc{\del}       {\partial}              % bra state
\nc{\bra}       {\langle}               % bra state
\nc{\ket}       {\rangle}               % ket state
\nc{\bras}[1]   {\langle #1|}           % bra state
\nc{\kets}[1]   {|#1\rangle}            % ket state
\nc{\mapleft}[1]{           % something under arrow
	\smash{\mathop{\,          %
			\hbox to 1.5cm{\rightarrowfill}\, }\limits_{#1}}}
\nc{\nn}      {\\\nonumber} \nc{\vs}      {\vspace{-0.275cm}}
\nc{\fra}    {\frac{1}{2}}
\nc{\mb}        {\mathbf}

\usepackage[colorlinks,linkcolor=blue,anchorcolor=teal,citecolor=red]{hyperref}

\begin{document}
	
	\preprint{}
	\title{Bayesian Inference of Dense-Matter Equations of State from Small-Radius Compact Stars with Twin-Star Scenarios } 
	
	\author{Xieyuan Dong}
	\affiliation{School of Physics, Nankai University, Tianjin 300071,  China}
    
	\author{Jinniu Hu}
	\email{hujinniu@nankai.edu.cn}
	\affiliation{School of Physics, Nankai University, Tianjin 300071,  China}
	\affiliation{Shenzhen Research Institute of Nankai University, Shenzhen 518083, China}
    
	\author{Ying Zhang}
    \email{yzhangjcnp@tju.edu.cn}
	\affiliation{Department of Physics, Faculty of Science, Tianjin University, Tianjin 300072, China}
    
	\author{Hong Shen}
	\affiliation{School of Physics, Nankai University, Tianjin 300071,  China}
    
	\date{\today}
    
   \begin{abstract}
   	We investigate dense-matter equations of state (EOSs) within a Bayesian framework, with particular emphasis on whether recent small-radius compact-star candidates can be accommodated in a twin-star scenario. For the hadronic sector, we adopt a meta-modeling EOS constrained by the NICER mass--radius measurements of PSR J0030$+$0451, PSR J0437$-$4715, PSR J0614$-$3329, and the massive pulsar PSR J0740$+$6620. The hadronic inference indicates that PSR J0614$-$3329 favors a somewhat softer EOS than the other two \(\sim1.4\,M_\odot\) pulsars, while the \(\sim2\,M_\odot\) constraint prevents the EOS from becoming too soft. We then introduce a strong first-order phase transition through a constant-speed-of-sound quark-matter segment. Using HESS J1731$-$347 and XTE J1814$-$338 to constrain the phase-transition parameters, we find a preferred transition density of \(n_\mathrm{t}\sim2.7\text{--}2.8\,n_0\), a sizable energy-density jump of \(600\text{--}700\) MeV, and a relatively large post-transition sound speed of \(c_s^2/c^2\sim0.85\). Such a phase transition generates a disconnected hybrid branch with radii of about \(6\text{--}7\) km at masses around \(1.2\text{--}1.4\,M_\odot\), and strongly suppresses the dimensionless tidal deformability relative to the purely hadronic branch. This pronounced change in tidal deformability is a characteristic signature of the twin-star mechanism and may provide an important observational tool for identifying phase transitions in neutron-star matter in future multimessenger measurements. These results show that small-radius compact stars can provide direct constraints on both the strength of a first-order phase transition and the stiffness of the post-transition phase in dense matter.
   	\end{abstract}
	
	%\pacs{21.10.Dr,  21.60.Jz,  21.80.+a}% PACS, the Physics and Astronomy
	%Use showkeys class option if keyword
	%display desired
	
	\maketitle
	%--------------------------------
\section{Introduction}\label{sec1}

Theoretical approaches to neutron-star (NS) structure are largely organized around constructing the equation of state (EOS) of cold dense matter and predicting the associated mass--radius ($M$--$R$) relation. For $\beta$-equilibrated NS ($npe\mu$) matter, two broad classes of methods have been extensively developed. On the one hand, {\it{ab initio}} many-body calculations based on realistic nucleon--nucleon interactions, such as Brueckner--Hartree--Fock (BHF) \cite{jaminon1989effective, hu2017nuclear}, relativistic Dirac--Brueckner--Hartree--Fock (DBHF) \cite{wang2020properties}, and variational approaches \cite{akmal1998equation}, aim to solve the many-body problem with controlled approximations. On the other hand, energy-density-functional theories with effective interactions, such as Skyrme--Hartree--Fock (SHF) \cite{stone2003nuclear, wang2024extended, duan2024new}, Gogny--Hartree--Fock (GHF) \cite{bender2003self, gonzalez2018new, vinas2021unified}, relativistic mean-field (RMF) \cite{zhang2018massive, hu2020effects, huang2022investigation, huang2024hadronic}, and relativistic Hartree--Fock (RHF) models \cite{sun2008neutron, zhu2016delta, liu2018nuclear}, provide flexible EOSs calibrated to finite nuclei and empirical nuclear matter properties. These frameworks are particularly useful for exploring the EOS systematics and for efficient stellar structure calculations across wide parameter ranges. In addition, parametrized EOS representations, for example, spectral or exponential expansions \cite{lindblom2010spectral, boyle2020parametrized} or meta-modeling based on polynomial expansions around saturation \cite{margueron2018equation1, margueron2018equation2, zhang2018combined, cai2025novel}, offer compact interfaces between theory and observations. Such parametrizations have become valuable tools for constraining saturation properties of symmetric nuclear matter and the density dependence of the symmetry energy using NS observations and heavy-ion collision data \cite{xie2021bayesian}.

With the rapid growth of NS observations, the field has increasingly adopted data-driven strategies that infer the EOS from macroscopic observables, i.e., masses, radii, and tidal deformabilities, rather than deriving it solely from a microscopic nuclear many-body model. This inversion perspective enables a direct mapping from measured NS properties to EOS parameters, while propagating observational uncertainties in a controlled manner. Recent progress has been driven by modern statistical and machine-learning techniques, including Bayesian inference \cite{huang2024constraining,  huang2025constraining, xie2020bayesian, malik2022bayesian, carvalho2023decoding, zhou2023bayesian, li2024bayesian}, support vector machines \cite{ferreira2021unveiling}, deep neural networks \cite{fujimoto2018methodology, zhou2023nonparametric, han2023nonparametric, zhou2024first, guo2024insights}, and so on. In particular, Bayesian inference combines prior information with observational likelihoods \cite{clark2007evidence, ozel2010astrophysical} to deliver posterior distributions for EOS parameters, thereby quantifying uncertainties and correlations in a transparent way. Machine-learning approaches provide a complementary route by learning complex, nonlinear mappings in high-dimensional parameter spaces, and can serve as emulators or surrogate models to accelerate inference.

Dense-matter EOSs constructed from conventional nuclear-physics inputs are broadly consistent with several well-measured NSs. Examples include PSR J0740$+$6620, with \(M=2.073^{+0.069}_{-0.069}\,M_\odot\) and \(R=12.49^{+1.28}_{-0.88}\)~km \cite{salmi2024radius}, PSR J0030$+$0451, with \(M=1.4^{+0.13}_{-0.12}\,M_\odot\) and \(R=11.71^{+0.88}_{-0.83}\)~km \cite{vinciguerra2024updated}, and PSR J0437$-$4715, with \(M=1.418\pm0.037\,M_\odot\) and \(R=11.36^{+0.95}_{-0.63}\)~km \cite{choudhury2024nicer}. More recently, several analyses have reported smaller-radius candidates, including PSR J0614$-$3329, with \(M=1.44^{+0.06}_{-0.07}\,M_\odot\) and \(R=10.29^{+1.01}_{-0.86}\)~km \cite{mauviard2025nicer} from NICER, the compact objects HESS J1731$-$347, with \(M=0.77^{+0.20}_{-0.17}\,M_\odot\) and \(R=10.4^{+0.86}_{-0.78}\)~km \cite{doroshenko2022strangely}, and XTE J1814$-$338, with \(M=1.21^{+0.05}_{-0.05}\,M_\odot\) and \(R=7.0^{+0.4}_{-0.4}\)~km \cite{kini2024constraining}. While PSR J0614$-$3329 may still be accommodated within purely hadronic descriptions, the extreme compactness inferred for XTE J1814$-$338 and the low-mass nature of HESS J1731$-$347 pose a challenge to many conventional EOS constructions, and motivate additional physical mechanisms that can produce substantial softening at intermediate densities followed by re-stiffening at higher densities.

A variety of scenarios has been proposed to account for these compact configurations. Yang {\it{et al.}} suggested a mixed population of strange stars and mirror dark matter to explain such observations \cite{yang2025strange,yang2025xte}. Alternative explanations include hybrid-star interpretations \cite{laskos2025xte, zhang2025slow}, the appearance of additional hadronic degrees of freedom such as \(K^{-}\) and \(\bar{K}^{0}\) condensates \cite{veselsky2025simultaneous}, speed-of-sound based phenomenology leading to a twin-star interpretation \cite{zhou2025hidden}, core--corona decomposition models \cite{zollner2025core}, and dark-matter admixed strange stars \cite{lopes2025macroscopic}. Among these, a strong first-order phase transition can naturally generate a ``twin-star'' (third-family) branch, offering a unified explanation for small radii while remaining compatible with the existence of $\sim 2\,M_\odot$ pulsars.

The concept of a ``third family'' of compact stars was originally discussed by Gerlach \cite{gerlach1968equation}, and the constant-speed-of-sound (CSS) parametrization introduced by Alford and Han provides a minimal yet versatile description of the post-transition stiffness \cite{alford2013generic}. Building on this framework, the combination of a hadronic meta-modeling baseline with a CSS high-density phase has been widely used to explore hybrid and twin-star properties \cite{zhang2025impact, grundler2025bayesian}. Recent studies have further investigated the twin-star parameter space \cite{christian2024first}, formation channels in full general relativity \cite{naseri2024exploring}, extensions to dark-matter admixed systems \cite{lu2025ultimate}, and observable imprints in postmerger gravitational-wave signals \cite{blacker2025comprehensive}, as well as the possibility of quark--quark phase transitions in hybrid quark stars \cite{yang2026hybrid}.

In this work, we employ Bayesian inference with a parametrized EOS to infer dense-matter properties from current $M$--$R$ information, with particular emphasis on small-radius compact stars. To assess whether these compact objects can be accommodated within a strong phase-transition scenario, we adopt the twin-star framework and combine a hadronic meta-modeling baseline with a CSS high-density phase. For the hadronic baseline we adopt the meta-modeling approach \cite{margueron2018equation1, margueron2018equation2,dong2025equation}, which parameterizes nuclear matter around saturation in terms of empirical or semi-empirical coefficients and yields an analytical EOS that is well suited for large-scale parameter scans. This flexibility enables a consistent incorporation of diverse theoretical and experimental constraints and makes meta-modeling particularly effective for multimessenger statistical analyses (e.g., masses, radii, and tidal deformabilities) \cite{zhang2018combined, zhang2019implications, zhang2020constraints, zhang2020gw190814, zhang2021impact}. Nevertheless, purely hadronic meta-modeling extrapolations at high density may lack microscopic guidance and therefore require additional physical conditions, such as causality and thermodynamic stability \cite{dong2025equation}.

This paper is organized as follows. Section~\ref{sec2} introduces the theoretical framework, including the meta-modeling EOS, the CSS parametrization, the stellar-structure equations, and the Bayesian methodology. In Sec.~\ref{sec3} we first constrain the hadronic EOS using NICER $M$--$R$ posteriors, with particular emphasis on the recent PSR J0614$-$3329 measurement. We then assess whether the CSS phase transition can accommodate XTE J1814$-$338 and HESS J1731$-$347, and present the resulting constraints on the post-transition stiffness as well as the implications for the $M$--$R$ relation, tidal deformabilities, and high-density diagnostics such as the sound speed and trace anomaly. Section~\ref{sec4} summarizes our findings and discusses future investigations.

\section{Theoretical framework}\label{sec2}

To model the interiors of compact stars in scenarios that may involve a strong first-order phase transition, we adopt a two-phase EOS construction consisting of a hadronic segment and a deconfined quark-matter segment. The hadronic phase is described by the meta-modeling EOS, which parametrizes cold $\beta$-equilibrated $npe\mu$ matter in terms of empirical nuclear-matter coefficients around saturation. The high-density phase is described by the CSS parametrization, providing a minimal yet versatile description of quark-matter thermodynamics suitable for Bayesian inference.

We employ the meta-modeling framework following Ref.~\cite{dong2025equation}, with foundational developments in Refs.~\cite{margueron2018equation1,margueron2018equation2} and combined experimental constraints discussed in Ref.~\cite{zhang2018combined}. The NS core is treated as uniform, cold $npe\mu$ matter, whose total energy density is decomposed as $\epsilon=\epsilon_b+\epsilon_l$, where $\epsilon_b$ and $\epsilon_l$ denote the baryonic and leptonic contributions, respectively.

The nuclear energy per baryon is expanded in the isospin asymmetry $\delta=(n_n-n_p)/n_b$ around symmetric nuclear matter (SNM),
\begin{equation}\label{eq:EA}
	\frac{E}{A}(n_b,\delta)=e_{\rm sat}(n_b)+e_{\rm sym}(n_b)\,\delta^2+\mathcal{O}(\delta^4),
\end{equation}
where $n_b=n_n+n_p$ is the baryon number density and $n_0=0.155\pm0.005~\mathrm{fm}^{-3}$ is the saturation density. Introducing the dimensionless variable $x=(n_b-n_0)/(3n_0)$, the SNM and symmetry-energy contributions are parametrized as
\begin{align}\label{eq:esat_esym}
	e_{\rm sat}(x) &= E_{\rm sat} + \frac{1}{2}K_{\rm sat}\,x^2 + \frac{1}{6}Q_{\rm sat}\,x^3, \nonumber\\
	e_{\rm sym}(x) &= E_{\rm sym} + L_{\rm sym}\,x + \frac{1}{2}K_{\rm sym}\,x^2 + \frac{1}{6}Q_{\rm sym}\,x^3,
\end{align}
where $\{E_{\rm sat},\,K_{\rm sat},\,Q_{\rm sat},\,E_{\rm sym},\,L_{\rm sym},\,K_{\rm sym},\,Q_{\rm sym}\}$ are the nuclear empirical parameters at saturation density and their higher-order density derivatives.

The baryonic energy density is then
\begin{equation}\label{eq:epsilon_b}
	\epsilon_b(x,\delta)=n_0(3x+1)\left[m_Nc^2+\frac{E}{A}(x,\delta)\right],
\end{equation}
with nucleon rest mass $m_N=939~\mathrm{MeV}/c^2$.

Leptons ($l=e,\mu$) are assumed as a zero-temperature, non-interacting relativistic Fermi gas,
\begin{equation}\label{eq:epsilon_l}
	\epsilon_l=\frac{m_l^4c^5}{8\pi^2\hbar^3}\left[t\sqrt{1+t^2}(1+2t^2)-\ln\!\left(t+\sqrt{1+t^2}\right)\right],
\end{equation}
where $t=\hbar k_F/(m_lc)$ and $k_F=(3\pi^2n_l)^{1/3}$.

The composition is fixed by $\beta$ equilibrium and charge neutrality:
$\mu_n-\mu_p=\mu_e=\mu_\mu$ and $n_p=n_e+n_\mu$.
Within the parabolic approximation, one has $\mu_n-\mu_p\simeq 4\delta\,e_{\rm sym}(n_b)$, while for leptons
$\mu_l=\sqrt{\hbar^2k_F^2c^2+m_l^2c^4}$.
For a given $(n_b,\delta)$, this yields the lepton density
\begin{equation}\label{eq:n_l}
	n_l=\frac{\left[16\delta^2e_{\rm sym}^2(n_b)-m_l^2c^4\right]^{3/2}}{3\pi^2\hbar^3c^3},
\end{equation}
and the proton density $n_p=(1-\delta)n_b/2$. The function $\delta(n_b)$ is obtained by solving the charge-neutrality condition self-consistently.

The pressure is computed from standard thermodynamic relations:
$p_b=n_b^2\,\partial(\epsilon_b/n_b)/\partial n_b$ for the baryons and $p_l=n_l\mu_l-\epsilon_l$ for each lepton species. The total EOS for the NS core is then $p=p_b+p_l$ and $\epsilon=\epsilon_b+\epsilon_l$.

For the quark phase, we adopt the CSS parametrization introduced in Ref.~\cite{alford2013generic}. The EOS is specified in the form $\epsilon(p)$ as
\begin{equation}
	\epsilon(p)=
	\begin{cases}
		\epsilon_{\mathrm{HM}}(p), & p < p_{\mathrm{t}}, \\[6pt]
		\epsilon_{\mathrm{HM}}(p_{\mathrm{t}}) + \Delta\epsilon + c_{\mathrm{QM}}^{-2}\,(p - p_{\mathrm{t}}), & p \ge p_{\mathrm{t}},
	\end{cases}
	\label{eq:css_eos}
\end{equation}
where $p_{\mathrm{t}}$ is the transition pressure from hadronic matter (HM) to quark matter (QM), $\Delta\epsilon$ is the discontinuity in energy density at the transition, and $c_{\mathrm{QM}}$ is the (constant) speed of sound in the quark phase.

In this work, we focus on the disconnected hybrid star (third-family) scenario, corresponding to a sufficiently strong first-order phase transition that generates a separate stable branch in the $M$--$R$ relation. Following Refs.~\cite{alford2013generic, grundler2025bayesian}, we restrict attention to parameter sets satisfying the Seidov stability criterion \cite{seidov1971stability},
\begin{equation}
	\frac{\Delta\epsilon}{\epsilon_{\mathrm{t}}}
	\ge \frac{1}{2} + \frac{3}{2}\,\frac{p_{\mathrm{t}}}{\epsilon_{\mathrm{t}}},
	\label{eq:seidov}
\end{equation}
where $\epsilon_{\mathrm{t}}\equiv \epsilon_{\mathrm{HM}}(p_{\mathrm{t}})$ is the hadronic energy density evaluated at the transition pressure. This condition ensures that the onset of the density jump does not immediately destabilize the stellar configuration, thereby allowing the existence of a disconnected hybrid-star branch.

Given an EOS $p(\epsilon)$, the structure of a nonrotating, spherically symmetric NS in general relativity is obtained by integrating the Tolman--Oppenheimer--Volkoff (TOV) equations \cite{tolman1939static,oppenheimer1939massive},
\begin{equation}\label{TOV}
	\begin{split}
		\frac{\mathrm{d} p(r)}{\mathrm{d} r}
		&=- \frac{Gm(r)\epsilon(r)}{c^2r^2}
		\left[1+\frac{p(r)}{\epsilon(r)}\right]
		\left[1+\frac{4{\pi}r^3p(r)}{c^2m(r)}\right]
		\left[1-\frac{2Gm(r)}{c^2r}\right]^{-1}, \\
		\frac{\mathrm{d} m(r)}{\mathrm{d} r}
		&=\frac{4 \pi r^2 \epsilon(r)}{c^2},
	\end{split}
\end{equation}
where $p(r)$, $\epsilon(r)$, and $m(r)$ denote the pressure, energy density, and enclosed mass at radius $r$, respectively.

Tidal deformability quantifies the quadrupolar response of a NS to an external tidal field \cite{hinderer2010tidal,read2013matter}. The dimensionless tidal deformability is
\begin{equation}
	\Lambda=\frac{2}{3}k_2 C^{-5},
\end{equation}
where $C=GM/(Rc^2)$ is the compactness and $k_2$ the dimensionless quadrupolar Love number,
\begin{equation}
	\begin{aligned}
		k_{2}= & \frac{8 C^{5}}{5}(1-2 C)^{2}\left[2-y_{R}+2 C\left(y_{R}-1\right)\right] \\
		& \times \left\{2 C\left[6-3 y_{R}+3 C\left(5 y_{R}-8\right)\right]\right. \\
		& +4 C^{3}\left[13-11 y_{R}+C\left(3 y_{R}-2\right)+2 C^{2}\left(1+y_{R}\right)\right] \\
		& \left. +3(1-2 C)^{2}\left[2-y_{R}+2 C\left(y_{R}-1\right) \ln (1-2 C)\right]\right\}^{-1},
	\end{aligned}
\end{equation}
with $y_R\equiv y(R)$. The function $y(r)$ is obtained by solving
\begin{equation}
	r \frac{\mathrm{d} y(r)}{\mathrm{d} r}+y^{2}(r)+y(r)\,F(r)+r^{2} Q(r)=0,
\end{equation}
subject to $y(0)=2$, simultaneously with the TOV system. The auxiliary functions are
\begin{equation}
	F(r)=\left[1-\frac{4 \pi r^{2} G}{c^{4}}\big(\epsilon(r)-p(r)\big)\right]\left(1-\frac{2 G m(r)}{r c^{2}}\right)^{-1},
\end{equation}
\begin{equation}
	\begin{aligned}
		r^{2} Q(r)= & \frac{4 \pi r^{2} G}{c^{4}}\left[5 \epsilon(r)+9 p(r)+\frac{\epsilon(r)+p(r)}{\partial p(r) \big/ \partial \epsilon(r)}\right] \left(1-\frac{2 G m(r)}{r c^{2}}\right)^{-1} \\
		& -6\left(1-\frac{2 G m(r)}{r c^{2}}\right)^{-1} \\
		& -\frac{4 G^{2} m^{2}(r)}{r^{2} c^{4}}\left(1+\frac{4 \pi r^{3} p(r)}{m(r) c^{2}}\right)^{2}\left(1-\frac{2 G m(r)}{r c^{2}}\right)^{-2}.
	\end{aligned}
\end{equation}

The parametrized EOS depends on a set of nuclear-matter coefficients that control the isoscalar and isovector behavior around saturation and the extrapolation to supranuclear densities, namely
$E_\mathrm{sat}$, $K_\mathrm{sat}$, $Q_\mathrm{sat}$, $E_\mathrm{sym}$, $L_\mathrm{sym}$, $K_\mathrm{sym}$, and $Q_\mathrm{sym}$.
Among them, the binding energy, incompressibility, and symmetry energy, $E_\mathrm{sat}$, $K_\mathrm{sat}$, and $E_\mathrm{sym}$ are relatively well constrained by finite-nuclei data and empirical systematics, whereas $Q_\mathrm{sat}$ and the isovector parameters $L_\mathrm{sym}$, $K_\mathrm{sym}$, and $Q_\mathrm{sym}$ remain substantially uncertain. We therefore combine astrophysical data with empirical priors and physical constraints to infer posterior distributions for the EOS parameters.

Let $\boldsymbol{\vartheta}$ denote the full parameter vector of the EOS model. Given observational data $\mathcal{D}$ (e.g., $M$--$R$ posteriors and additional astrophysical constraints), the likelihood is $p(\mathcal{D}\mid\boldsymbol{\vartheta})$. Bayes' theorem yields the posterior distribution
\begin{equation}
	p(\boldsymbol{\vartheta} \mid \mathcal{D})
	= \frac{p(\mathcal{D} \mid \boldsymbol{\vartheta})\, p(\boldsymbol{\vartheta})}{p(\mathcal{D})},
\end{equation}
where $p(\boldsymbol{\vartheta})$ is the prior, and the evidence
$p(\mathcal{D})=\int p(\mathcal{D}\mid\boldsymbol{\vartheta})p(\boldsymbol{\vartheta})\,\mathrm{d}\boldsymbol{\vartheta}$
acts as a normalization constant.

\section{Results and discussion}\label{sec3}

The discussion in this section proceeds in two stages. We first consider the purely hadronic hypothesis within the meta-modeling EOS and infer the higher-order empirical parameters from current $M$--$R$ observations. We then introduce a strong first-order phase transition through the CSS parametrization and investigate whether the twin-star mechanism can accommodate the compact objects HESS J1731$-$347 and XTE J1814$-$338, as well as what this implies for the high-density EOS.

We begin with the hadronic baseline. Using the NICER $M$--$R$ constraints of three pulsars with masses around \(1.4\,M_\odot\), namely PSR J0030$+$0451, PSR J0437$-$4715, and PSR J0614$-$3329, together with the massive pulsar PSR J0740$+$6620, we perform Bayesian inference for the meta-modeling EOS. We first perform separate inferences for the three lower-mass pulsars and then a joint inference including all four pulsars. Each inference consists of 300,000 iterations of Bayesian analysis. In this part of the analysis, the poorly constrained higher-order empirical parameters \((Q_{\rm sat},\,L_{\rm sym},\,K_{\rm sym},\,Q_{\rm sym})\) are sampled, while the better constrained saturation properties are fixed by empirical nuclear information. The observational constraints are thus mapped directly onto posterior distributions of the hadronic EOS parameters and, subsequently, onto the corresponding family of stellar configurations.

% fig1
\begin{figure}[htbp]
	\centering
	\includegraphics[width=0.6\textwidth]{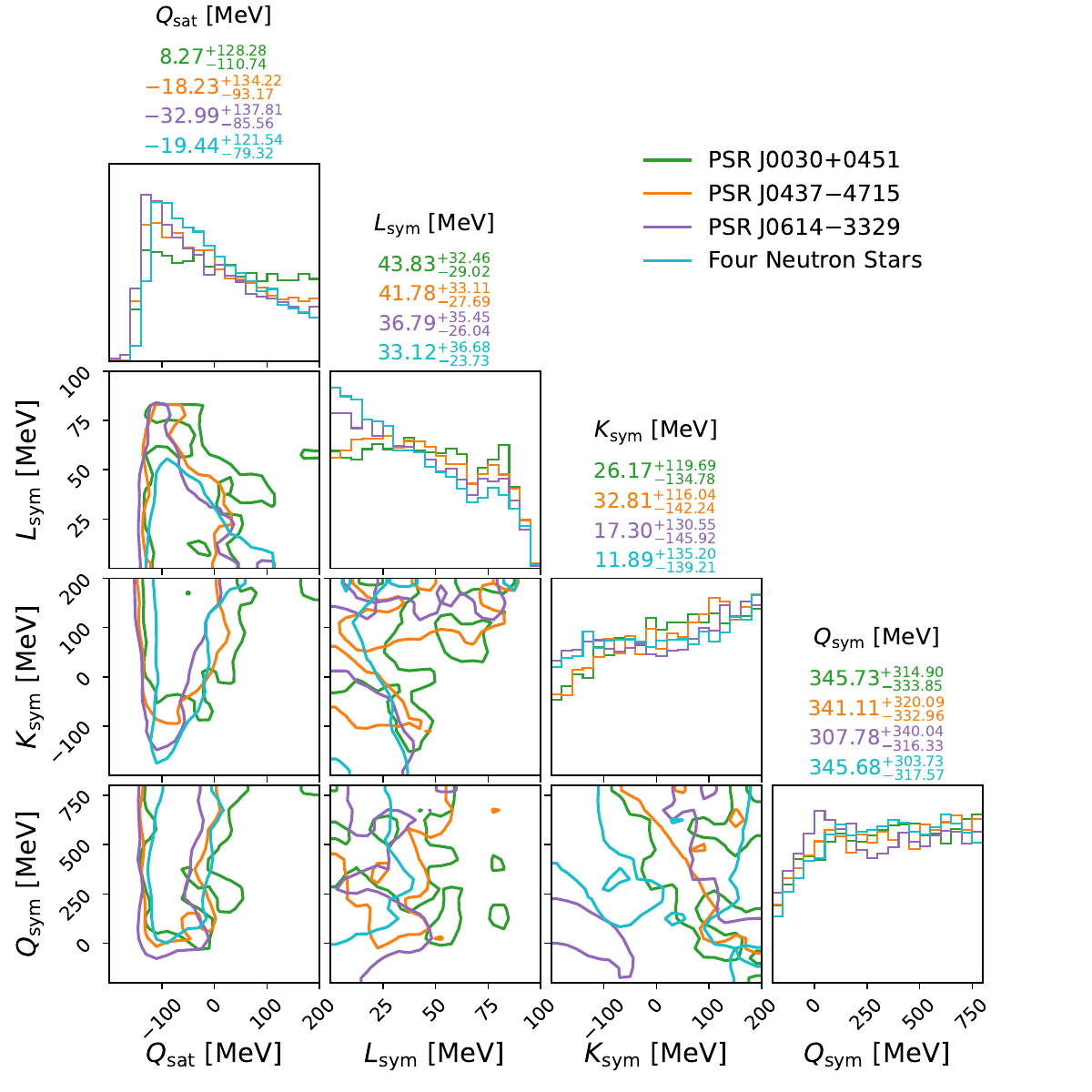}
	\caption{Marginalized and joint posterior distributions of the nuclear matter parameters \(Q_{\rm sat}\), \(L_{\rm sym}\), \(K_{\rm sym}\), and \(Q_{\rm sym}\) inferred from PSR J0030$+$0451, PSR J0437$-$4715, PSR J0614$-$3329, and the joint analysis of four NSs. The contours correspond to the 1$\sigma$ credible interval.}
	\label{fig1}
\end{figure}

The resulting marginalized posterior distributions are shown in Fig.~\ref{fig1}, and the corresponding parameter constraints are summarized in Table~\ref{tab1}. Relative to the adopted priors, the observational information leads to visible posterior concentration for all four sampled parameters. The constraints inferred from PSR J0030$+$0451 and PSR J0437$-$4715 are broadly similar, indicating that these two \(\sim1.4\,M_\odot\) pulsars favor comparable hadronic EOS properties. By contrast, PSR J0614$-$3329 systematically shifts the posterior toward lower median values of
\(Q_{\rm sat}\), \(L_{\rm sym}\), \(K_{\rm sym}\), and \(Q_{\rm sym}\).

% tab1
\begingroup
\setlength{\tabcolsep}{6pt}
\renewcommand{\arraystretch}{1}
\begin{table}[htbp]
	\centering
    \caption{Prior and posterior distributions (1$\sigma$ CIs) for the higher-order isoscalar and isovector parameters.}
	\begin{tabular}{cccccccc}
		\hline \hline
		& $Q_{\text{sat}}$ [MeV] & $L_{\text{sym}}$ [MeV] & $K_{\text{sym}}$ [MeV] & $Q_{\text{sym}}$ [MeV] \\
		\hline
		Prior & $[-200,\ 200]$ & $[0,\ 100]$ & $[-200,\ 200]$ & $[-200,\ 800]$ \\
		PSR J0030$+$0451 & $8.27_{-110.74}^{+128.28}$ & $43.83_{-29.02}^{+32.46}$ & $26.17_{-134.78}^{+119.69}$ & $345.73_{-333.85}^{+314.90}$ \\
		PSR J0437$-$4715 & $-18.23_{-93.17}^{+134.22}$ & $41.78_{-27.69}^{+33.11}$ & $32.81_{-142.24}^{+116.04}$ & $341.11_{-332.96}^{+320.09}$ \\
		PSR J0614$-$3329 & $-32.99_{-85.56}^{+137.81}$ & $36.79_{-26.04}^{+35.45}$ & $17.30_{-145.92}^{+130.55}$ & $307.78_{-316.33}^{+340.04}$ \\
		Four NSs         & $-19.44_{-79.32}^{+121.54}$ & $33.12_{-23.73}^{+36.68}$ & $11.89_{-139.21}^{+135.20}$ & $345.68_{-317.57}^{+303.73}$ \\
		\hline\hline
	\end{tabular}
	\label{tab1}
\end{table}
\endgroup  

This tendency is quantitatively reflected in Table~\ref{tab1}. For example, the median value of \(L_{\rm sym}\) decreases from \(43.83\) MeV for PSR J0030$+$0451 and \(41.78\) MeV for PSR J0437$-$4715 to \(36.79\) MeV for PSR J0614$-$3329. The same trend appears in the other three parameters as well. This indicates that PSR J0614$-$3329 favors a somewhat softer hadronic EOS in the density region relevant for canonical-mass NSs. Once all four NSs are included simultaneously, the allowed parameter region becomes narrower overall, showing that the combined requirement of describing both the \(\sim1.4\,M_\odot\) pulsars and the \(\sim2\,M_\odot\) star PSR J0740$+$6620 provides a stronger global constraint.

\begin{figure}[htbp]
	\centering
	\includegraphics[width=0.6\textwidth]{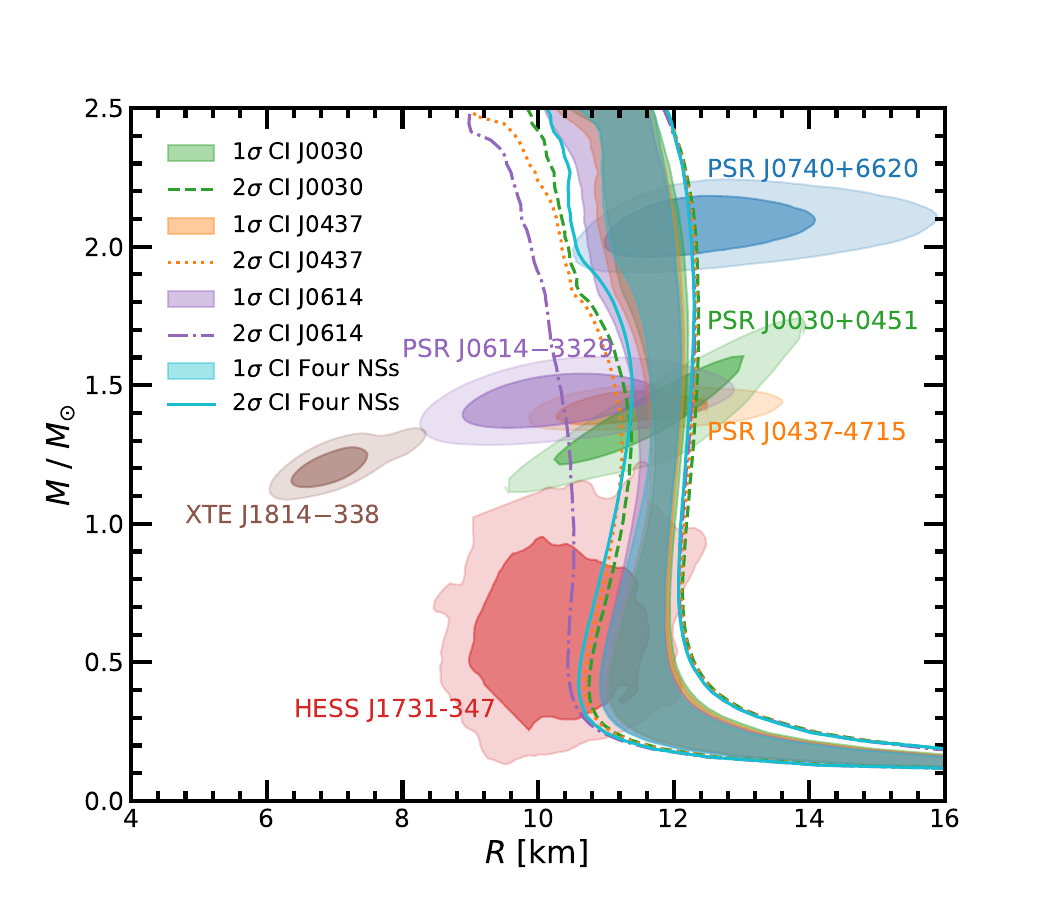}
	\caption{Hadronic $M$--$R$ relations inferred from the individual analyses of PSR J0030$+$0451, PSR J0437$-$4715, and PSR J0614$-$3329, together with the joint analysis of four NSs and the corresponding observational constraints.}
	\label{fig2}
\end{figure}

The corresponding stellar consequence is shown in Fig.~\ref{fig2}, which displays the hadronic $M$--$R$ relations reconstructed from the inferred parameter distributions. The shaded regions denote the 1$\sigma$ and 2$\sigma$ credible intervals (CIs). For the three single-pulsar analyses, the reconstructed $M$--$R$ bands all pass through the observational region of the respective source. The branch inferred from PSR J0614$-$3329 is shifted toward smaller radii at intermediate masses, which is the direct stellar-structure manifestation of the softer parameter trend seen in Fig.~\ref{fig1}. In the joint analysis, the allowed $M$--$R$ band becomes visibly tighter and remains compatible with the observational regions of PSR J0030$+$0451, PSR J0437$-$4715, PSR J0614$-$3329, and PSR J0740$+$6620. Within the purely hadronic meta-modeling framework, the current data can therefore still be accommodated self-consistently, although the preference of PSR J0614$-$3329 for smaller radii already points toward increasing tension with a purely hadronic interpretation if more compact objects are included. We note that the present hadronic inference is primarily driven by low- and intermediate-mass NSs, so the highest-density part of the purely hadronic EOS is not the main focus here. A fully consistent treatment including an explicit causality constraint on the hadronic sector has been considered in Ref. \cite{dong2025equation}.

% fig2

For completeness, we briefly compare the present hadronic inference with our previous meta-modeling study, Ref.~\cite{dong2025equation}. The present analysis uses the same meta-modeling framework, but incorporates updated observational information, in particular the recent small-radius source PSR J0614$-$3329 together with the massive pulsar PSR J0740$+$6620. Compared with Ref.~\cite{dong2025equation}, the inferred hadronic EOS is therefore shifted toward somewhat smaller radii at intermediate masses, while still satisfying the \(\sim2\,M_\odot\) constraint. In this sense, the difference between the two studies mainly reflects the impact of the newer compact-star data.

\begin{figure}[htbp]
	\centering
	\includegraphics[width=1\textwidth]{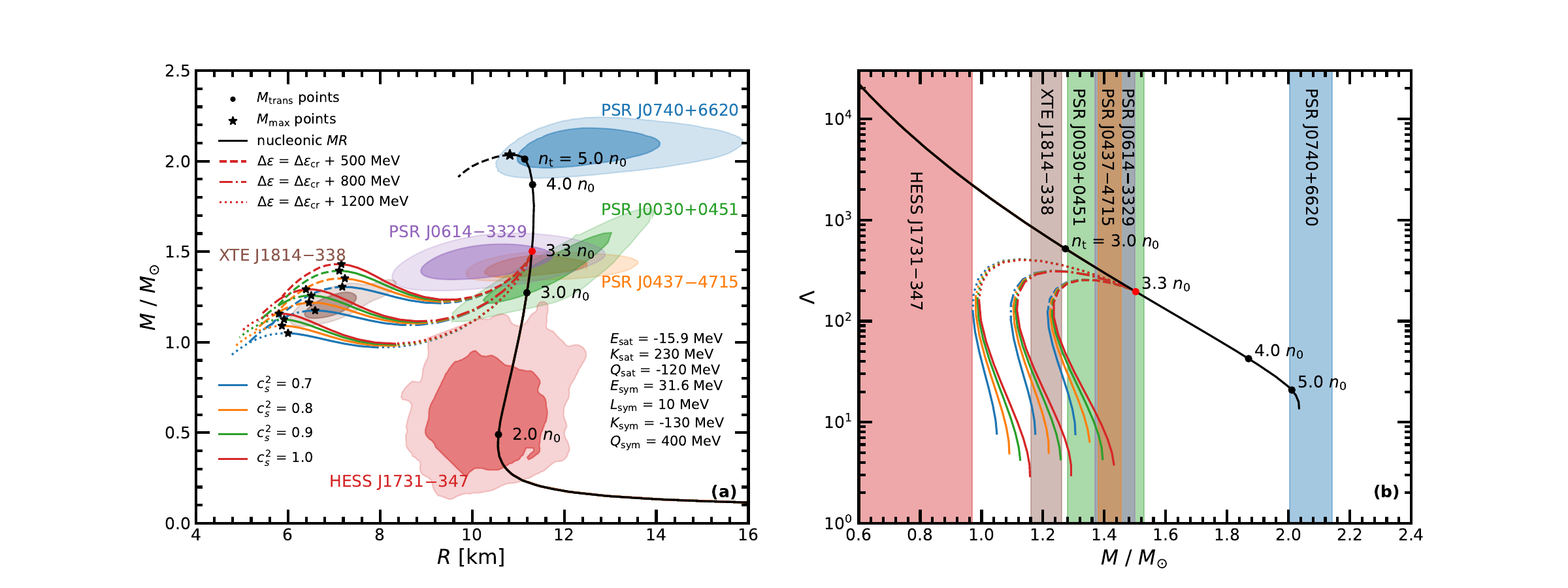}
	\caption{ \(M\)--\(R\) curves (a) and dimensionless  tidal deformability (b) for twin stars derived from the CSS model with specified nuclear matter parameters. The black solid line represents the \(M\)--\(R\)  curve for pure hadronic matter.}
	\label{fig3}
\end{figure}

To examine whether this trend can persist in the presence of a strong phase transition, we next introduce the CSS model. For illustration, we first select a representative hadronic EOS from the meta-modeling posterior,
$Q_{\rm sat}=-120~\mathrm{MeV},~L_{\rm sym}=10~\mathrm{MeV},
~K_{\rm sym}=-130~\mathrm{MeV}$,
and
$Q_{\rm sym}=400~\mathrm{MeV}$,
and use it as a fiducial baseline. This choice is not intended to replace the full posterior propagation. Rather, it allows us to visualize how the transition strength and the post-transition stiffness control the emergence of a disconnected hybrid branch. The maximum NS mass from this hadronic EOS as $2.03M_\odot$ with $R_\mathrm{max}=10.81$ km, when the center density is around $5n_0$, can accommodate the constraint from the massive pulsar PSR J0740+6620 and approaches $2M_\odot$.

With this fiducial baseline, we first fix the transition density to \(n_\mathrm{t}=3.3\ n_0\), for which the hadronic branch reaches a stellar mass of about \(1.4\,M_\odot\), and then vary both the energy-density jump and the quark-phase sound speed. Specifically, we consider \(\Delta\epsilon=\Delta\epsilon_{\rm cr}+500\), \(800\), and \(1200~\mathrm{MeV}\), together with \(c_s^2/c^2=0.7\), \(0.8\), \(0.9\), and \(1.0\).
The corresponding $M$--$R$ relations and dimensionless tidal deformabilities are shown in Fig.~\ref{fig3}. Panel (a) shows that once the phase transition is sufficiently strong, a disconnected hybrid branch appears at substantially smaller radii than the purely hadronic sequence. As a representative example, the post-transition branch can reach radii of only about \(6\text{--}7\) km at masses around \(1.2\text{--}1.4\,M_\odot\), whereas the hadronic branch in the same mass range remains near \(R\sim 11\) km. Panel (b) shows that the same mechanism also suppresses the tidal response. Around \(M\sim1.3\,M_\odot\), the purely hadronic sequence gives \(\Lambda\) values of the order of several hundreds, while the hybrid branch can reduce them to the order of \(10^2\) or less. Fig.~\ref{fig3} therefore illustrates the basic role of the phase transition: it generates a much more compact branch and simultaneously lowers the tidal deformability over the same mass range.

We next examine whether the compact sources themselves can constrain the phase-transition parameters. We therefore apply the same two-phase construction to HESS J1731$-$347 with Bayesian inference. For a given set of CSS parameters, the post-transition branch contains a minimum-mass configuration that marks the onset of the disconnected hybrid sequence, and we use the mass and radius of this point to construct the likelihood for HESS J1731$-$347. In addition, we perform a joint analysis combining HESS J1731$-$347 with XTE J1814$-$338, in order to test whether a single twin-star framework can simultaneously accommodate both compact objects.

% fig4
\begin{figure}[htbp]
	\centering
	\includegraphics[width=0.6\textwidth]{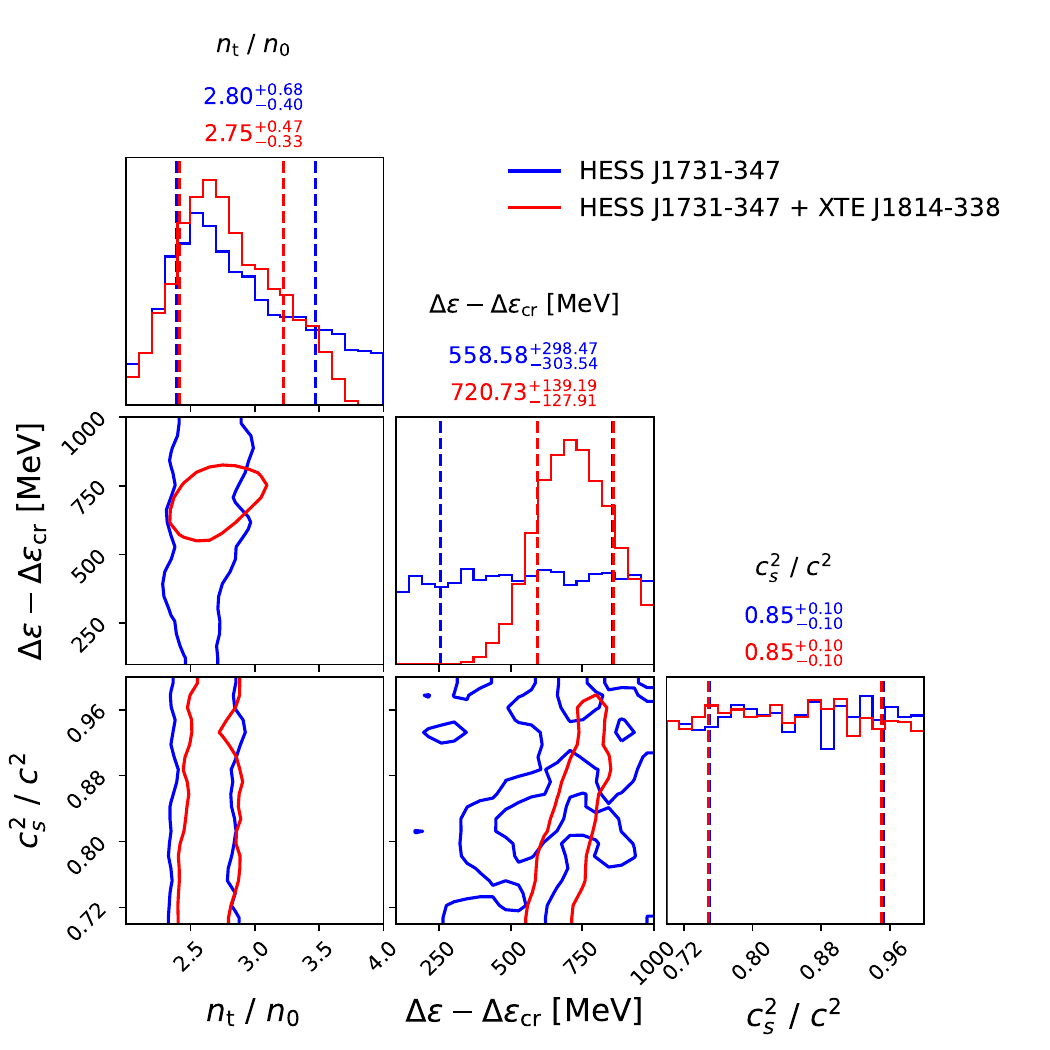}
	
	\caption{Marginalized and joint posterior distributions of the CSS parameters inferred from HESS J1731$-$347 alone and from the joint analysis of HESS J1731$-$347 and XTE J1814$-$338.}
	\label{fig4}
\end{figure}

The posterior distributions for the CSS parameters are displayed in Fig.~\ref{fig4}, and the corresponding 1$\sigma$ CIs are listed in Table~\ref{tab2}. HESS J1731$-$347 alone already favors a transition density of a few times saturation,
\(n_{\mathrm{t}}=2.80^{+0.68}_{-0.40}\,n_0\), together with a sizable transition strength, \(\Delta\epsilon-\Delta\epsilon_{\rm cr}=588.58^{+298.47}_{-303.54}~\mathrm{MeV}\), and a stiff post-transition phase, \(c_s^2/c^2=0.85^{+0.10}_{-0.10}\).
After XTE J1814$-$338 is included, the transition density becomes modestly tighter, \(n_{\mathrm{t}}=2.75^{+0.47}_{-0.33}\,n_0\), while the transition strength shifts upward and becomes much more concentrated, \(\Delta\epsilon-\Delta\epsilon_{\rm cr}=720.73^{+139.19}_{-127.91}~\mathrm{MeV}\). By contrast, the posterior of \(c_s^2/c^2\) remains comparatively broad. These posterior trends suggest that simultaneously accommodating HESS J1731$-$347 and XTE J1814$-$338 requires a strong transition together with a sufficiently stiff post-transition phase, while the current data remain less sensitive to the precise value of the sound speed once it is already large.

% tab2
\begingroup
\setlength{\tabcolsep}{6pt}
\renewcommand{\arraystretch}{1}
\begin{table}[htbp]
	\centering
	\caption{Prior and posterior distributions (1$\sigma$ CIs) on the CSS parameters}
	\begin{tabular}{ccccccc}
		\hline \hline
		& $n_{\mathrm{t}}$ / $n_0$ & $\Delta \epsilon - \Delta \epsilon_{\rm cr}$ [MeV] & $c_s^2$ / $c^2$ \\
		\hline
		Prior & $[2.0,\ 4.0]$ & $[100,\ 1000]$ & $[0.7,\ 1.0]$ \\
		HESS J1731$-$347                   & $2.80_{-0.40}^{+0.68}$ & $588.58_{-303.54}^{+298.47}$ & $0.85_{-0.10}^{+0.10}$ \\
		HESS J1731$-$347 + XTE J1814$-$338 & $2.75_{-0.33}^{+0.47}$ & $720.73_{-127.91}^{+139.19}$ & $0.85_{-0.10}^{+0.10}$ \\
		\hline\hline
	\end{tabular}
	\label{tab2}
\end{table}
\endgroup  

\begin{figure}[htbp]
    \centering
    \includegraphics[width=1\textwidth]{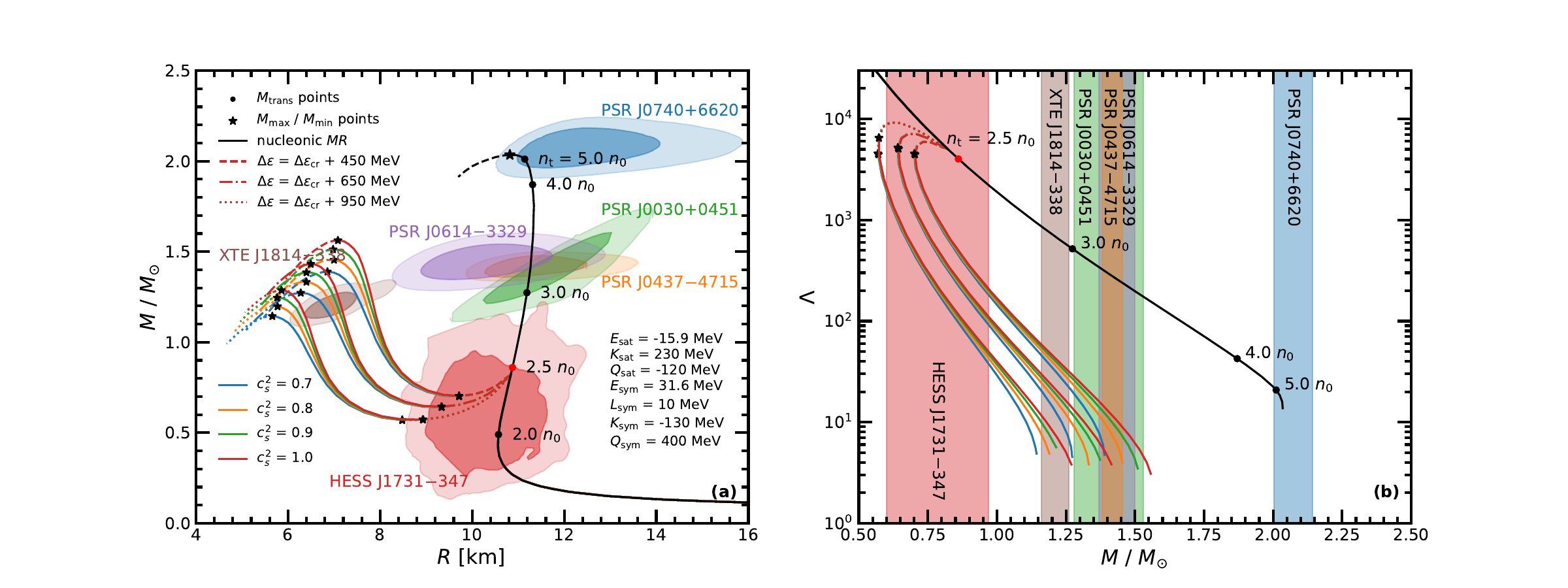}
    \caption{Same as Fig.~\ref{fig3}, but with the phase transition density set to \(2.5\,n_0\). The \(\Delta\epsilon\) values are \(\Delta\epsilon_{\mathrm{cr}}+450\), \(650\), and \(950\,\mathrm{MeV}\).}
    \label{fig5}
\end{figure}

A more concrete picture emerges when we construct representative twin-star sequences for a lower transition density, $n_{\mathrm{t}}=2.5\,n_0$, which lies within the inferred range and is more relevant for producing low-mass compact configurations. We consider \(\Delta\epsilon=\Delta\epsilon_{\rm cr}+450\), \(650\), and \(950~\mathrm{MeV}\),
together with \(c_s^2/c^2=0.7\), \(0.8\), \(0.9\), and \(1.0\). The resulting $M$--$R$ relations and dimensionless tidal deformabilities are shown in Fig.~\ref{fig5}, while the maximum masses and corresponding radii of the post-transition branch are summarized in Table~\ref{tab3}. Compared with the $n_{\mathrm{t}}=3.3\,n_0$case, the disconnected hybrid branch now appears more prominently in the low- and intermediate-mass region. As a representative example, compact configurations with \(R\sim6\text{--}7\) km can already occur at \(M\sim1.2\text{--}1.4\,M_\odot\), making this setup more directly relevant for HESS J1731$-$347 and XTE J1814$-$338. Table~\ref{tab3} further shows how the phase-transition parameters control the branch. At fixed \(c_s^2/c^2=0.8\), increasing
\(\Delta\epsilon-\Delta\epsilon_{\rm cr}\) from \(450\) to \(950~\mathrm{MeV}\)
reduces the maximum mass from \(1.455\,M_\odot\) to \(1.197\,M_\odot\), while the corresponding radius decreases from \(6.997\) km to \(5.771\) km. On the other hand, at fixed transition strength, increasing \(c_s^2\) systematically raises \(M_{\rm max}\), showing that the sound speed mainly determines how much stability can be recovered after the transition.

% fig5

% tab3
\begin{table}[htbp]
\centering
\caption{Maximum mass $M_{\rm max}$ and corresponding radius $R_{\rm max}$ for different values of $\Delta\epsilon - \Delta \epsilon_{\rm cr}$ (in MeV) and squared sound speed $c_s^2$.}
\label{tab3}
\begin{tabular}{ccccc}
\hline
$\Delta\epsilon - \Delta \epsilon_{\rm cr}$ [MeV] & $c_s^2$ / $c^2$ & $M_{\rm max}$ /$M_\odot$ & $R_{\rm max}$ [km] \\
\hline
\multirow{4}{*}{450} 
 & 0.7 & 1.389 & 6.855 \\
 & 0.8 & 1.455 & 6.997 \\
 & 0.9 & 1.512 & 6.978 \\
 & 1.0 & 1.562 & 7.083 \\
\hline
\multirow{4}{*}{650} 
 & 0.7 & 1.272 & 6.275 \\
 & 0.8 & 1.333 & 6.399 \\
 & 0.9 & 1.385 & 6.399 \\
 & 1.0 & 1.431 & 6.494 \\
\hline
\multirow{4}{*}{950} 
 & 0.7 & 1.144 & 5.657 \\
 & 0.8 & 1.197 & 5.771 \\
 & 0.9 & 1.245 & 5.762 \\
 & 1.0 & 1.287 & 5.857 \\
\hline
\end{tabular}
\end{table}

Fig.~\ref{fig5}(b) presents the corresponding dimensionless tidal deformability \(\Lambda\) as a function of mass. Compared with the purely hadronic branch, the twin-star solutions occupy a distinctly lower-\(\Lambda\) region over the same mass interval. As a representative example, for masses around \(1.0\text{--}1.3\,M_\odot\), the hybrid branches typically give \(\Lambda\) values from a few tens to a few hundreds, whereas the purely hadronic sequence remains at several hundreds or above. Therefore, the transition not only reduces the stellar radius, but also suppresses the tidal response very efficiently.

The effect on the tidal response is quantified in Table~\ref{tab4}. For the representative case \(\Delta\epsilon=\Delta\epsilon_{\rm cr}+650~\mathrm{MeV}\) and \(c_s^2/c^2=0.8\),
the table compares \(\Lambda(M)\) for the hybrid branch and the purely hadronic branch. At \(M=1.0\,M_\odot\), for instance, the deformability decreases from \(\Lambda_{\rm HB}=1928.60\) to \(\Lambda_{\rm MM}=82.05\),
corresponding to a ratio of only \(0.04\). Across the mass interval from \(0.7\) to \(1.3\,M_\odot\), the ratio
\(\Lambda_{\rm HB}/\Lambda_{\rm PHB}\)
falls from \(0.12\) to \(0.02\), demonstrating that the twin-star branch suppresses the dimensionless tidal deformability by roughly one to two orders of magnitude. This strong suppression of \(\Lambda\) is a direct consequence of the much smaller radii of the post-transition branch and provides an additional observational discriminator beyond the $M$--$R$ relation alone.

% tab4
\begin{table}[htbp]
\centering
\caption{Tidal deformability $\Lambda$ for NSs in the hybrid branch (HB) and the purely hadronic branch (PHB) at different masses $M$, calculated with $\Delta \epsilon = \Delta \epsilon_{\rm cr} + 650\,\rm MeV$ and $c_s^2 = 0.8$.}
\label{tab4}
\begin{tabular}{cccc}
\hline
$M$ ($M_\odot$) & $\Lambda_{\rm HB}$ & $\Lambda_{\rm PHB}$ & $\Lambda_{\rm HB} / \Lambda_{\rm PHB}$ \\
\hline
0.7 & 1274.00 & 11029.27 & 0.12 \\
0.8 & 425.62  & 5847.84  & 0.07 \\
0.9 & 177.59  & 3234.22  & 0.05 \\
1.0 & 82.05   & 1928.60  & 0.04 \\
1.1 & 38.96   & 1177.73  & 0.03 \\
1.2 & 18.24   & 735.13   & 0.02 \\
1.3 & 7.21    & 467.60   & 0.02 \\
\hline
\end{tabular}
\end{table}

\begin{figure}[htbp]
	\centering
	\includegraphics[width=0.6\textwidth]{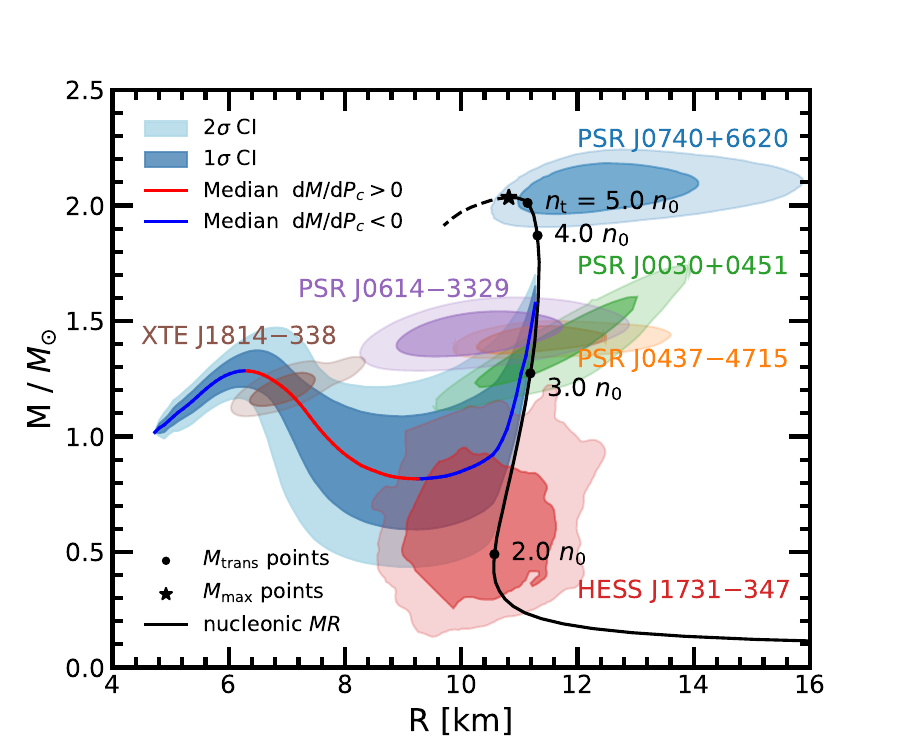}
	\caption{Reconstructed post-transition $M$--$R$ relations from the joint posterior of HESS J1731$-$347 and XTE J1814$-$338. The shaded regions denote the 1$\sigma$ and 2$\sigma$ CIs, while the median stable and unstable branches are shown separately.}
	\label{fig6}
\end{figure}

Returning to the full posterior reconstruction for the joint HESS J1731$-$347 + XTE J1814$-$338 analysis, Fig.~\ref{fig6} shows the reconstructed post-transition $M$--$R$ bands. The stable hybrid branch is concentrated in the small-radius region, mainly around radii of \(6\text{--}8\) km and masses of \(1.0\text{--}1.4\,M_\odot\), clearly separated from the purely hadronic sequence. In particular, the 1$\sigma$ credible interval already overlaps the \(1\sigma\) observational region of XTE J1814$-$338, and the 2$\sigma$ CI covers most of its \(2\sigma\) range. The agreement with HESS J1731$-$347 is more limited: only part of the posterior family intersects its observational region, and much of that overlap lies near the unstable branch. This comparison suggests that the joint posterior appears to be primarily shaped by the requirement of producing a very compact stable branch for XTE J1814$-$338, while HESS J1731$-$347 remains more difficult to accommodate simultaneously within the same construction.

% fig6

\begin{figure}[htbp]
	\centering
	\includegraphics[width=0.4\textwidth]{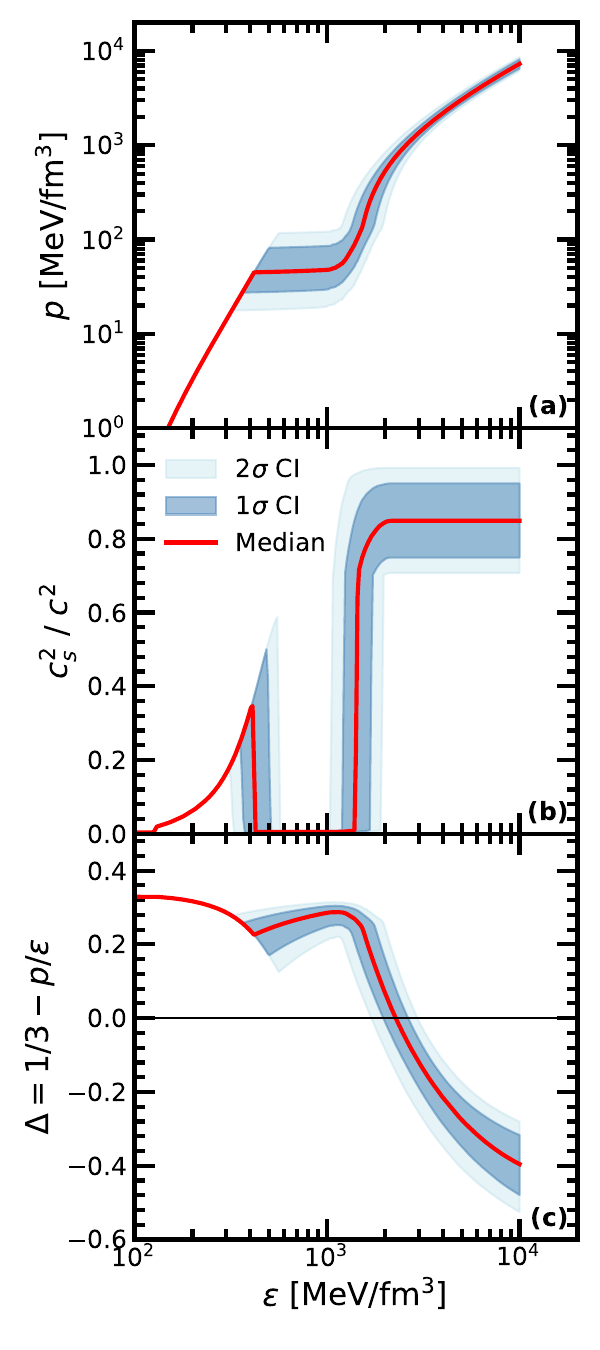}
	\caption{Posterior reconstructions of (a) the EOS \(p(\epsilon)\), (b) the squared sound speed \(c_s^2\), and (c) the trace anomaly \(\Delta=1/3-p/\epsilon\). The median and the 1$\sigma$ and 2$\sigma$ CI are shown in each panel.}
	\label{fig7}
\end{figure}

From the perspective of high-density diagnostics, Fig.~\ref{fig7} displays the posterior reconstructions of the EOS \(p(\epsilon)\), the squared sound speed \(c_s^2=\partial p/\partial\epsilon\), and the trace anomaly, defined here as \(\Delta\equiv 1/3-p/\epsilon\).
The reconstructed EOS band in panel (a) is already fairly localized, indicating that the combined compact-star constraints significantly restrict the allowed post-transition pressure at a given energy density. Panel (b) shows that the inferred squared sound speed remains high throughout the relevant post-transition interval; the median curve is around \(c_s^2/c^2\sim0.8\) in the central region, with the 1$\sigma$ CI also favoring values well above the conformal limit \(1/3\). Panel (c) complements this picture by showing that \(\Delta\) stays substantially above zero over the same density range, with a typical median value of order \(0.2\), implying a sizable deviation from conformality. The reconstructed diagnostics therefore point to a post-transition phase that is both stiff and significantly nonconformal over the density interval relevant to the disconnected hybrid branch.

% fig7

\section{Summary and outlook}\label{sec4}

In this work, we investigated dense-matter equations of state within a Bayesian framework, with particular emphasis on whether recent small-radius compact-star candidates can be accommodated in a twin-star scenario. Our analysis proceeded in two stages. We first inferred the higher-order empirical parameters of the meta-modeling EOS under the purely hadronic hypothesis using current NICER $M$--$R$ constraints. We then introduced a strong first-order phase transition through the CSS parametrization and examined its consequences for compact stars such as HESS J1731$-$347 and XTE J1814$-$338, as well as for the high-density behavior of the EOS.

At the hadronic level, the NICER constraints from PSR J0030$+$0451, PSR J0437$-$4715, and PSR J0614$-$3329, together with the massive pulsar PSR J0740$+$6620, already provide meaningful posterior constraints on the higher-order empirical parameters ($Q_{\rm sat}$, $L_{\rm sym}$, $K_{\rm sym}$, $Q_{\rm sym}$). In the joint analysis of the four NSs, we obtained, for example,
\(Q_{\rm sat}=-19.44^{+121.54}_{-79.32}\) MeV and \(L_{\rm sym}=33.12^{+36.68}_{-23.73}\) MeV,
while PSR J0614$-$3329 systematically favored lower median values than PSR J0030$+$0451 and PSR J0437$-$4715. The corresponding reconstructed $M$--$R$ relations show that the purely hadronic meta-modeling framework can still accommodate the present data self-consistently, although the smaller-radius preference of PSR J0614$-$3329 already points toward increasing tension if additional compact objects are included.

Motivated by this tension, we explored the effect of a strong first-order phase transition through the CSS model. The illustrative scans show that such a transition can generate a disconnected hybrid branch with radii of about \(6\text{--}7\) km at masses around \(1.2\text{--}1.4\,M_\odot\), while simultaneously suppressing the dimensionless tidal deformability by one to two orders of magnitude relative to the purely hadronic branch. This demonstrates that the twin-star mechanism provides a natural way to accommodate very compact configurations within a unified EOS framework.

We then used HESS J1731$-$347 and XTE J1814$-$338 to constrain the phase-transition parameters directly. For HESS J1731$-$347 alone, the posterior already favors a transition density of a few times saturation, $n_{\mathrm{t}}=2.80^{+0.68}_{-0.40}\,n_0$,
together with a sizable transition strength, \(\Delta\epsilon-\Delta\epsilon_{\rm cr}=588.58^{+298.47}_{-303.54}\) MeV.
When XTE J1814$-$338 is included, the posterior becomes more localized and shifts toward a stronger transition,
$n_{\mathrm{t}}=2.75^{+0.47}_{-0.33}\,n_0$ and \(\Delta\epsilon-\Delta\epsilon_{\rm cr}=720.73^{+139.19}_{-127.91}\) MeV,
while the preferred squared sound speed remains high, \(c_s^2/c^2 \simeq 0.85\).
These results indicate that the present compact-star data are already informative about how strong the phase transition must be, while remaining less sensitive to the precise value of the post-transition sound speed once the high-density phase is sufficiently stiff.

The full posterior reconstruction further clarifies the resulting astrophysical picture. The reconstructed post-transition $M$--$R$ branch is concentrated in the small-radius region and can naturally overlap the observational range of XTE J1814$-$338, whereas the agreement with HESS J1731$-$347 is more limited and often lies near the unstable branch. At the same time, the reconstructed high-density diagnostics show that the post-transition EOS is relatively well localized in the \(p(\epsilon)\) plane, that the squared sound speed remains well above the conformal limit over the relevant density interval, and that the trace anomaly
\(\Delta=1/3-p/\epsilon\) stays substantially different from zero. These features point to a post-transition phase that must stiffen rapidly enough to recover stability while remaining significantly nonconformal.

Several extensions of the present work are worth pursuing. First, it will be important to propagate the full posterior uncertainty of the hadronic meta-modeling EOS directly into the CSS inference, rather than relying on representative hadronic baselines for illustrative scans. Second, additional multimessenger information, especially gravitational-wave constraints on dimensionless tidal deformability and future improved radius measurements, will provide a more stringent test of whether the disconnected hybrid branch is truly favored over a purely hadronic interpretation. Third, it will be useful to go beyond the CSS parametrization and examine whether the present conclusions remain robust in more general descriptions of quark matter. Finally, the strong reduction of the dimensionless tidal deformability predicted on the post-transition branch suggests that future observations of low- and intermediate-mass compact stars may provide a particularly sensitive probe of strong first-order phase transitions in dense matter.

\section{Acknowledgments}
This work was supported  by the National Natural Science Foundation of China No. 12475149, and the Guangdong Basic and Applied Basic Research Foundation (Grant  No. 2024A1515010911). 

\section{Data Availability}
The  python code and data that support the findings of this article are openly available in zenodo \cite{zenodo}.

\bibliographystyle{apsrev4-2}
\bibliography{refer}

\end{document}